\documentclass[conference]{IEEEtran}
\usepackage{multirow,graphicx,framed,subfigure}
\usepackage[hyphens]{url}

\begin{document}
\title{Analyzing Social and Stylometric Features to Identify Spear phishing Emails}

\author{\IEEEauthorblockN{Prateek Dewan$^*$, Anand Kashyap$^\dagger$, Ponnurangam Kumaraguru$^*$}
\IEEEauthorblockA{$^*$Indraprastha Institute of Information Technology, Delhi\\
$^*$Cybersecurity Education and Research Centre (CERC), IIIT-Delhi\\
$^\dagger$Symantec Research Labs\\
Email: $^*$\{prateekd,pk\}@iiitd.ac.in, $^\dagger$anand\_kashyap@symantec.com}}



%

\maketitle

\begin{abstract}
Targeted social engineering attacks in the form of spear phishing emails, are often the main gimmick used by attackers to infiltrate organizational networks and implant state-of-the-art Advanced Persistent Threats (APTs). Spear phishing is a complex targeted attack in which, an attacker harvests information about the victim prior to the attack. This information is then used to create sophisticated, genuine-looking attack vectors, drawing the victim to compromise confidential information. What makes spear phishing different, and more powerful than normal phishing, is this contextual information about the victim. Online social media services can be one such source for gathering vital information about an individual. In this paper, we characterize and examine a true positive dataset of spear phishing, spam, and normal phishing emails from Symantec's enterprise email scanning service. We then present a model to detect spear phishing emails sent to employees of 14 international organizations, by using \emph{social} features extracted from 
LinkedIn. 
Our dataset consists of 4,742 targeted attack emails sent to 2,434 victims, and 9,353 non targeted attack emails sent to 5,912 non victims; and publicly available information from their LinkedIn profiles. We applied various machine learning algorithms to this labeled data, and achieved an overall maximum accuracy of 97.76\% in identifying spear phishing emails. We used a combination of \emph{social} features from LinkedIn profiles, and stylometric features extracted from email subjects, bodies, and attachments. However, we achieved a slightly better accuracy of 98.28\% without the \emph{social} features. Our analysis revealed that \emph{social} features extracted from LinkedIn do not help in identifying spear phishing emails. To the best of our knowledge, this is one of the first attempts to make use of a combination of stylometric features extracted from emails, and \emph{social} features extracted from an online social network to detect targeted spear phishing emails.

\end{abstract}

%

\IEEEpeerreviewmaketitle

\section{Introduction}

A new race of insidious threats called Advanced Persistent Threats (APTs) have joined the league of eCrime activities on the Internet, and caught a lot of organizations off guard in the fairly recent times. Critical infrastructures and the governments, corporations, and individuals supporting them are under attack by these increasingly sophisticated cyber threats. The goal of the attackers is to gain access to intellectual property, personally identifiable information, financial data, and targeted strategic information. This is not simple fraud or hacking, but intellectual property theft and infrastructure corruption on a grand scale~\cite{Daly2009}. APTs use multiple attack techniques and vectors that are conducted by stealth to avoid detection, so that hackers can retain control over target systems unnoticed for long periods of time. Interestingly, no matter how sophisticated these attack vectors may be, the most common ways of getting them inside an organization's network are social engineering attacks like phishing, and targeted spear phishing emails. There have been numerous reports of spear phishing attacks causing losses of millions of dollars in the recent past.~\footnote{\url{http://businesstech.co.za/news/internet/56731/south-africas-3-billion-phishing-bill/}}~\footnote{\url{http://www.scmagazine.com/stolen-certificates-used-to-deliver-trojans-in-spear-phishing-campaign/article/345626/}} Although there exist antivirus, and other similar protection software to mitigate such attacks, it is always better to stop such vectors at the entry level itself~\cite{kumaraguru2010teaching}. This requires sophisticated techniques to deter spear phishing attacks, and identify malicious emails at a very early stage itself.

\begin{figure}[!h]
     \begin{center}
\fbox{\includegraphics[scale=0.61]{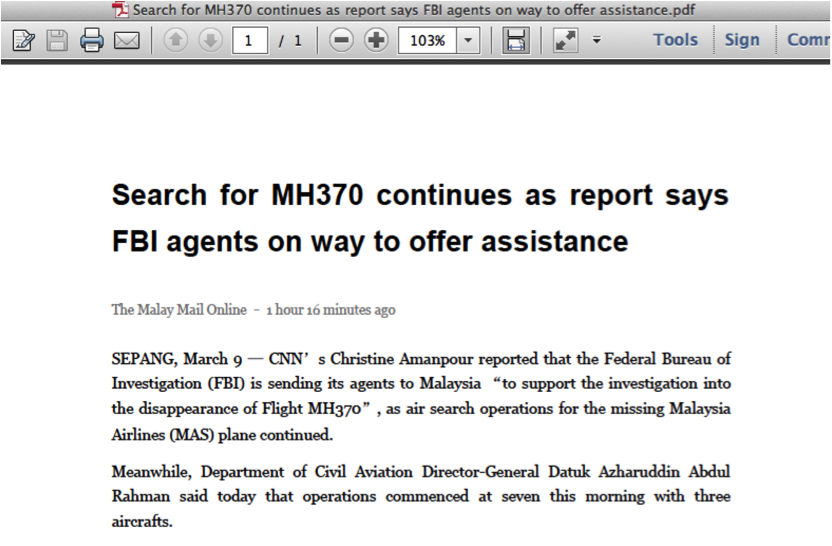}}
    \end{center}
    \caption{Example of a malicious PDF attachment sent via a spear phishing email. The PDF attachment was said to contain information about the missing Malaysian Airlines Flight 370.
     }
   \label{fig:mh370attach}

\end{figure}

In this research paper, we focus on identifying such spear phishing emails, wherein the attacker targets an individual or company, instead of anyone in general. Spear phishing emails ususally contain victim-specific context instead of general content. Since it is targeted, a spear phishing attack looks much more realistic, and thus, harder to detect~\cite{Jakobsson2006}. A typical spear phishing attack can broadly be broken down into two phases. In the first phase, the attacker tries to gather as much information about the victim as possible, in order to craft a scenario which looks realistic, is believable for the victim, and makes it very easy for the attacker to attain the victim's trust. In the second phase, the attacker makes use of this gained trust, and draws the victim into giving out sensitive / confidential information like a user name, password, bank account details, credit card details, etc. The attacker can also exploit the victim's trust by infecting the victim's system, through luring them into downloading and opening malicious attachments~\cite{Jakobsson2006}. While spear phishing may be a timeworn technique, it continues to be effective even in today's Web 2.0 landscape. A very recent example of such a spear phishing attack was reported by FireEye. Here, attackers exploited the news of the disappearance of Malaysian Airlines Flight MH370, to lure government officials across the world into opening malicious attachments (Figure~\ref{fig:mh370attach}) sent to them over email~\cite{fireeye:2014}. In 2011, security firm RSA suffered a breach via a targeted attack; analysis revealed that the compromise began with the opening of a spear phishing email.~\footnote{\url{http://blogs.rsa.com/rivner/anatomy-of-an-attack/}} That same year, email service provider Epsilon also fell prey to a spear phishing attack that caused the organization to lose an estimated US\$4 billion.~\footnote{\url{http://www.net-security.org/secworld.php?id=10966}} These examples indicate that spear phishing has been, and continues to be one of the biggest forms of eCrime over the past few years, especially in terms of the monetary losses incurred.

Spear phishing was first studied as context aware phishing by Jakobsson et al. in 2005~\cite{Jakobsson2005}. A couple of years later, Jagatic et al. performed a controlled experiment and found that the number of victims who fell for context aware phishing / spear phishing is 4.5 times the number of victims who fell for general phishing~\cite{Jagatic2007}. This work was preliminary proof that spear phishing attacks have a much higher success rate than normal phishing attacks. It also highlighted that, what separates a regular phishing attack from a spear phishing attack is the additional information / context. Online social media services like LinkedIn, which provide rich professional information about individuals, can be one such source for extracting contextual information, which can be used against a victim. Recently, the FBI also warned that spear phishing emails typically contain accurate information about victims often obtained from data posted on social networking sites, blogs or other websites.~\footnote{\url{http://www.computerweekly.com/news/2240187487/FBI-warns-of-increased-spear-phishing-attacks}} In this work, we investigate if publicly available information from LinkedIn can help in differentiating a spear phishing from a non spear phishing email received by an individual. 
We attained a dataset of true positive targeted spear phishing emails, and a dataset of a mixture of non targeted, spam and phishing attack emails from the Symantec's enterprise email scanning service, which is deployed at multiple international organizations around the world. To conduct the analysis at the organizational level, we extracted the most frequently occurring domains from the \emph{``to"} fields of these emails, and filtered out 14 most targeted organizations, where the first name and last name could be derived from the email address. 
Our final dataset consisted of 4,742 spear phishing emails sent to 2,434 unique employees, and 9,353 non targeted spam / phishing emails to 5,914 unique employees. For a more exhaustive analysis, we also used a random sample of 6,601 benign emails from the Enron email dataset~\cite{Cohen2009} sent to 1,240 unique employees with LinkedIn profiles.

We applied 4 classification algorithms, and were able to achieve a maximum accuracy of 97.04\% for classifying spear phishing, and non spear phishing emails using a combination of \emph{email} features, and \emph{social} features. However, without the \emph{social} features, we were able to achieve a slightly higher accuracy of 98.28\% for classifying these emails. We then looked at the most informative features, and found that \emph{email} features performed better than \emph{social} features at differentiating targeted spear phishing emails from non targeted spam / phishing emails, and benign Enron emails. To the best of our knowledge, this is the first attempt at making use of a social media profile of a user to distinguish targeted spear phishing emails from non targeted attack emails, received by her. 
Having found that \emph{social} features extracted from LinkedIn profiles do not help in distinguishing spear phishing and non spear phishing emails, our results encourage to explore other social media services like Facebook, and Twitter. Such studies can be particularly helpful in mitigating APTs, and reducing the chances of attacks to an organization at the entry level itself. 

The rest of the paper is arranged as follows. Section~\ref{sec:bg} discuss the related work, Section~\ref{sec:dcm} describes our email and LinkedIn profile datasets, and the data collection methodology. The analysis and results are described in Section~\ref{sec:ar}. We conclude our findings, and discuss the limitations, contributions, and scope for future work in Section~\ref{sec:conclusion}.

\section{Background and Related work} \label{sec:bg}

The concept of targeted phishing was first introduced in 2005 as \emph{social phishing} or \emph{context-aware phishing}~\cite{Jakobsson2005}. Authors of this work argued that if the attacker can infer or manipulate the context of the victim before the attack, this context can be used to make the victim volunteer the target information. This theory was followed up with an experiment where Jagatic et al. harvested freely available acquaintance data of a group of Indiana University students, by crawling social networking websites like Facebook, LinkedIn, MySpace, Orkut etc.~\cite{Jagatic2007}. This contextual information was used to launch an actual (but harmless) phishing attack targeting students between the age group of 18-24 years. Their results indicated about 4.5 times increase in the number of students who fell for the attack which made use of contextual information, than the generic phishing attack. However, authors of this work do not provide details of the kind and amount of information they were able to gather from social media websites about the victims.

\paragraph{Who falls for phish}

Dhamija et al. provided the first empirical evidence about which malicious strategies are successful at deceiving general users~\cite{Dhamija2006}. Kumaraguru et al. conducted a series of studies and experiments on creating and evaluating techniques for teaching people not to fall for phish~\cite{kumaraguru2009school,kumaraguru2007protecting,kumaraguru2010teaching}. Lee studied data from Symantec's enterprise email scanning service, and calculated the odds ratio of being attacked for these users, based on their area of work. The results of this work indicated that users with subjects \emph{``Social studies"}, and \emph{``Eastern, Asiatic, African, American and Australasian Languages, Literature and Related Subjects"} were both positively correlated with targeted attacks at more than 95\% confidence~\cite{Lee2012}. Sheng et al. conducted an online survey with 1,001 participants to study who is more susceptible to phishing based on demographics. Their results indicated that women are more susceptible than men to phishing, and participants between the ages 18 and 25 are more susceptible to phishing than other age groups~\cite{Sheng2010}. In similar work, Halevi et al. found a strong correlation between gender and response to a prize phishing email. They also found that neuroticism is the factor most correlated to responding to the email. Interestingly, authors detected no correlation between the participants' estimate of being vulnerable to phishing attacks and actually being phished. This suggests susceptibility to phishing is not due to lack of awareness of the phishing risks, and that real-time response to phishing is hard to predict in advance by online users~\cite{Halevi2013}. 

\paragraph{Phishing email detection techniques}

To keep this work focused, we concentrate only on techniques proposed for detecting phishing emails; we do not cover all the techniques used for detecting phishing URLs or phishing websites in general. Abu-Nimeh et al.~\cite{Abu-Nimeh2007} studied the performance of different classifiers used in text mining such as Logistic regression, classification and regression trees, Bayesian additive regression trees, Support Vector Machines, Random forests, and Neural networks. Their dataset consisted of a public collection of about 1,700 phishing mails, and 1,700 legitimate mails from private mailboxes. They focused on richness of word to classify phishing email based on 43 keywords. The features represent the frequency of ``bag-of-words" that appear in phishing and legitimate emails. However, the ever-evolving techniques and language used in phishing emails might make it hard for this approach to be effective over a long period of time.

Various feature selection approaches have also been recently introduced to assist phishing detection. A lot of previous work~\cite{Abu-Nimeh2007,Chandrasekaran2006,Fette2007} has focused on email content in order to classify the emails as either benign or malicious. Chandrasekaran et al.~\cite{Chandrasekaran2006} presented an approach based on natural structural characteristics in emails. The features included number of words in the email, the vocabulary, the structure of the subject line, and the presence of 18 keywords. They tested on 400 data points which were divided into five sets with different type of feature selection. Their results were the best when more features were used to classify phishing emails using Support Vector Machine. Authors of this work proposed a rich set of stylometric features, but the dataset they used was very small as compared to a lot of other similar work. Fette et al.~\cite{Fette2007} on the other hand, considered 10 features which mostly examine URL and presence of JavaScript to flag emails as phishing. Nine features were extracted from the email and the last feature was obtained from WHOIS query. They followed a similar approach as Chandrasekaran et al. but using larger datasets, about 7,000 normal emails and 860 phishing emails. Their filter scored 97.6\% F-measure, false positive rate of 0.13\% and a false negative rate of 3.6\%. The heavy dependence on URL based features, however, makes this approach ineffective for detecting phishing emails which do not contain a URL, or are attachment based attacks, or ask the user to reply to the phishing email with potentially sensitive information. URL based features were also used by Chhabra et al. to detect phishing using short URLs~\cite{Chhabra2011}. Their work, however, was limited to only URLs, and did not cover phishing through emails. Islam and Abawajy~\cite{Islam2013} proposed a multi-tier phishing detection and filtering approach. They also proposed a method for extracting the features of phishing email based on weights of message content and message header. The results of their experiments showed that the proposed algorithm reduces the false positive problems substantially with lower complexity.

Behavior-based approaches have also been proposed by various researchers to determine phishing messages~\cite{Toolan2010,Zhang2007a}. Zhang et al.~\cite{Zhang2007a} worked on detecting abnormal mass mailing hosts in network layer by mining the traffic in session layer. Toolan et al.~\cite{Toolan2010} investigated 40 features that have been used in recent literature, and proposed behavioral features such as number of words in \emph{sender} field, total number of characters in \emph{sender} field, difference between sender's domain and reply-to domain, and difference between sender's domains and the email's modal domain, to classify ham, spam, and phishing emails. Ma et al.~\cite{Ma2009} attempted to identify phishing email based on hybrid features. They derived 7 features categorized into three classes, i.e. content features, orthographic features, and derived features, and applied 5 machine learning algorithms. Their results stated that Decision Trees worked best in identifying phishing emails. Hamid et al.~\cite{Hamid2013} proposed a hybrid feature selection approach based on combination of content and behaviour. Their approach mined attacker behavior based on email header, and achieved an accuracy of 94\% on a publicly available test corpus.

All of the aforementioned work concentrates on distinguishing phishing emails from legitimate ones, using various types of features extracted from email content, URLs, header information etc. To the best of our knowledge, there exists little work which focuses specifically on targeted spear phishing emails. Further, there exists no work which utilizes features from the social media profiles of the victim in order to distinguish an attack email from a legitimate one. In this work, we direct our focus on a very specific problem of distinguishing targeted spear phishing emails from general phishing, spam, and benign emails. Further, we apply \emph{social} features extracted from the LinkedIn profile of recipients of such emails to judge whether an email is a spear phishing email or not; which has never been attempted before, to the best of our knowledge. We performed our entire analysis on a real-world dataset derived from Symantec's enterprise email scanning service.

\section{Data collection methodology} \label{sec:dcm}

The dataset we used for the entire analysis, is a combination of two separate datasets, viz. a dataset of emails (combination of targeted attack and non targeted attack emails), and a dataset of LinkedIn profiles. We now explain both these datasets in detail. 

\subsection{Email dataset} \label{sec:emd}

\begin{table*}[!ht]
\begin{center}
    \begin{tabular}{l|l||l|l}
    \hline
    Spear phishing Attachment Name & \% & Spam / phishing Attachment Name        & \%  \\ \hline
    work.doc  & 3.46       & 100A\_0.txt          & 20.74 \\
More detail Chen Guangcheng.rar  & 3.01 & 100\_5X\_AB\_PA1\_MA-OCTET-STREAM\_\_form.html  & 9.02 \\
    ARMY\_600\_8\_105.zip  & 2.54     & ./attach/100\_4X\_AZ-D\_PA2\_\_FedEx=5FInvoice=5FN56=2D141.exe       & 4.19 \\
    Strategy\_Meeting.zip    & 1.58    & 100\_2X\_PM3\_EMS\_MA-OCTET=2DSTREAM\_\_apply.html                            & 2.66  \\
    20120404 H 24 year annual business plan 1 quarterly.zip & 1.33   & 100\_4X\_AZ-D\_PA2\_\_My=5Fsummer=5Fphotos=5Fin=5FEgypt=5F2011.exe            & 1.87  \\
    The extension of the measures against North Korea.zip   & 1.30 & ./attach/100\_2X\_PM2\_EMS\_MA-OCTET=2DSTREAM\_\_ACC01291731.rtf   & 1.40  \\
  Strategy\_Meeting\_120628.zip & 1.28  & 100\_5X\_AB\_PA1\_MH\_\_NothernrockUpdate.html               & 1.28  \\
     image.scr & 1.24 & ./attach/100\_2X\_PM2\_EMS\_MA-OCTET=2DSTREAM\_\_invoice.rtf                  & 1.15  \\
    Consolidation Schedule.doc     & 0.98  & 100\_6X\_AZ-D\_PA4\_\_US=2DCERT=20Operations=20Center=20Report=2DJan2012.exe  & 1.12  \\
    DARPA-BAA-11-65.zip                                     & 0.93     & 100\_4X\_AZ-D\_PA2\_\_I=27m=5Fwith=5Fmy=5Ffriends=5Fin=5FEgypt.exe            & 1.11  \\
    Head Office-Y drive.zip                                 & 0.93   & 100\_4X\_AZ-D\_PA2\_\_I=27m=5Fon=5Fthe=5FTurkish=5Fbeach=5F2012.exe           & 0.80  \\
    page 1-2.doc   & 0.90 & 100\_5X\_AB\_PA1\_MA-OCTET-STREAM\_\_Lloyds=R01TSB=R01-=R01Login=R01Form.html & 0.69  \\
    Aircraft Procurement Plan.zip      & 0.90   & 100\_6X\_AZ-D\_PA4\_\_Fidelity=20Investments=20Review=2Dfrom=2DJan2012.exe    & 0.68  \\
   Overview of Health Reform.doc                           & 0.74   & 100\_4X\_AZ-D\_PA2\_\_FedEx=5FInvoice=20=5FCopy=5FIDN12=2D374.exe     & 0.64  \\
   page 1-2.pdf  & 0.64  & 100\_4X\_AZ-D\_PA2\_\_my=5Fphoto=5Fin=5Fthe=5Fdominican=5Frepublic.exe        & 0.63  \\
     fisa.pdf  & 0.58   & 100\_2X\_PM4\_EMQ\_MH\_\_message.htm          & 0.60  \\
   urs.doc & 0.52    & /var/work0/attach/100\_4X\_AZ-D\_PA2\_\_document.exe    & 0.58  \\
   script.au3   & 0.50    & 100\_6X\_AZ-D\_PA4\_\_Information.exe          & 0.58  \\
    install\_reader10\_en\_air\_gtbd\_aih.zip    & 0.48   & /var/work0/attach/100\_4X\_AZ-D\_PA2\_\_Ticket.exe     & 0.57  \\
    dodi-3100-08.pdf   & 0.43    & 100\_4X\_AZ-D\_PA2\_\_Ticket.exe     & 0.57  \\ \hline
    \end{tabular}
\vspace{10pt}
\caption{Top 20 most frequently occurring attachment names, and their corresponding percentage share in our spear phishing and spam / phishing datasets. Attachment names in the spear phishing emails dataset look much more realistic and genuine as compared to attachment names in spam / phishing emails dataset.}
\label{tab:attach_names}
\end{center}
\end{table*}

Our email dataset consisted of a combination of targeted spear phishing emails, non targeted spam and phishing emails, and benign emails. We obtained the targeted spear phishing emails from Symantec's enterprise email scanning service. \emph{Symantec} collects data regarding targeted attacks that consist of emails with malicious attachments. These emails are identified from the vast majority of non-targeted malware by evidence of there being prior research and selection of the recipient, with the malware being of high sophistication and low copy number. The process by which Symantec's enterprise mail scanning service collects such malware has already been described elsewhere~\cite{Thonnard2012,Lee2013}. The corpus almost certainly omits some attacks, and most likely also includes some non-targeted attacks, but nevertheless it represents a large number of sophisticated targeted attacks compiled according to a consistent set of criteria which render it a very useful dataset to study.

The non targeted attack emails were also obtained from Symantec's email scanning service. These emails were marked as \emph{malicious}, and were a combination of malware, spam, and phishing. Both these datasets contained an enormously large number of emails received at hundreds of organizations around the world, where Symantec's email scanning services are being used. Before selecting a suitable sample for organization level analysis, we present an overview of this entire data. Table~\ref{tab:attach_names} shows the top 20 most frequently occurring attachment names in the complete spear phishing and spam / phishing datasets. We found distinct differences in the type of attachment names in these two datasets. While names in spear phishing emails looked fairly genuine and personalized, attachment names in spam / phishing emails were irrelevant, and long. It was also interesting to see that the attachment names associated with spear phishing emails were less repetitive than those associated with spam / phishing emails. As visible in Table~\ref{tab:attach_names}, the most commonly occurring attachment name in spear phishing emails was found in less than 3.5\% of all spear phishing emails, while in the case of spam / phishing emails, the most common attachment name was present in over 20\% of all spam / phishing emails. This behavior reflects that attachments in spear phishing emails are named more carefully, and with more effort to make them look genuine.

We also looked at the most frequently spread file types in spear phishing, spam, and phishing emails. Table~\ref{tab:attach_types} shows the top 15 most frequently occurring file types in both the spear phishing and spam / phishing email datasets. Not surprisingly, both these datasets had notable presence of executable (.exe, .bat, .com), and compressed (.rar, .zip, .7z) file types. In fact, most of the file types spread through such emails were among the most frequently used file types in general, too. Microsoft Word, Excel, PowerPoint, and PDF files were also amongst the most frequently spread files. It was, however, interesting to note that lesser percentage of targeted spear phishing emails contained executables than spam / phishing emails. This reflects that attackers prepare targeted attacks smartly as compared to spammers / phishers, and avoid using executables, which are more prone to suspicion.

\begin{table}[!h]
\begin{center}
    \begin{tabular}{p{2.7cm}|p{0.6cm}||p{3cm}|p{0.8cm}}
    \hline
    Spear phishing Attachment Type               & \%  & Spam / phishing Attachment Type               & \%     \\ \hline
    Zip archive data (zip)              & 19.59 & Windows Executable (exe)           & 38.39 \\
    PDF document (pdf)               & 13.73 & ASCII text (txt)                             & 21.73   \\
    Composite Document File     & 13.63 & Hypertext (html)                          & 18.08   \\
    Windows Executable (exe)     & 11.20 & Hypertext (htm)                           & 7.06   \\
    Rich Text Format data (rtf)    & 10.40 & Rich Text Format data (rtf)          & 3.04    \\
    RAR archive data (rar)            & 9.47  & PDF document (pdf)                     & 2.04    \\
    Screensaver (scr)                   & 5.06  & Zip archive data (zip)                    & 1.75    \\
    data (dat)                              & 3.00  &  Microsoft Word                            & 1.27    \\
    JPEG image data (jpg)           & 1.64  &  Screensaver (scr)                          &  1.14   \\
    CLASS                                   & 1.56  & Microsoft Excel (xls)                     & 0.81    \\
    Microsoft Word 2007+         & 1.15  & Program Info. file (pif)                   & 0.80    \\
    7-zip archive data (7z)         &1.12  & Dynamic-link Library (dll)             & 0.30     \\
    Shortcut (lnk)                      & 1.08  & Windows Batch file (.bat)                & 0.24     \\
    ASCII text (txt)                    & 0.80    & JavaScript (js)                               & 0.17     \\
    Dynamic-link Library (dll)   & 0.54    & Microsoft HTML Help (chm)         & 0.16     \\ \hline
    \end{tabular}
\vspace{5pt}
\caption{Top 15 most frequently occurring attachment types, and their corresponding percentage share in our spear phishing and spam / phishing datasets. Only 5 file types were common in the top 15 in these datasets.}
\label{tab:attach_types}
\end{center}
\end{table}

\begin{figure*}[!ht]
     \begin{center}
\fbox{\includegraphics[scale=0.45]{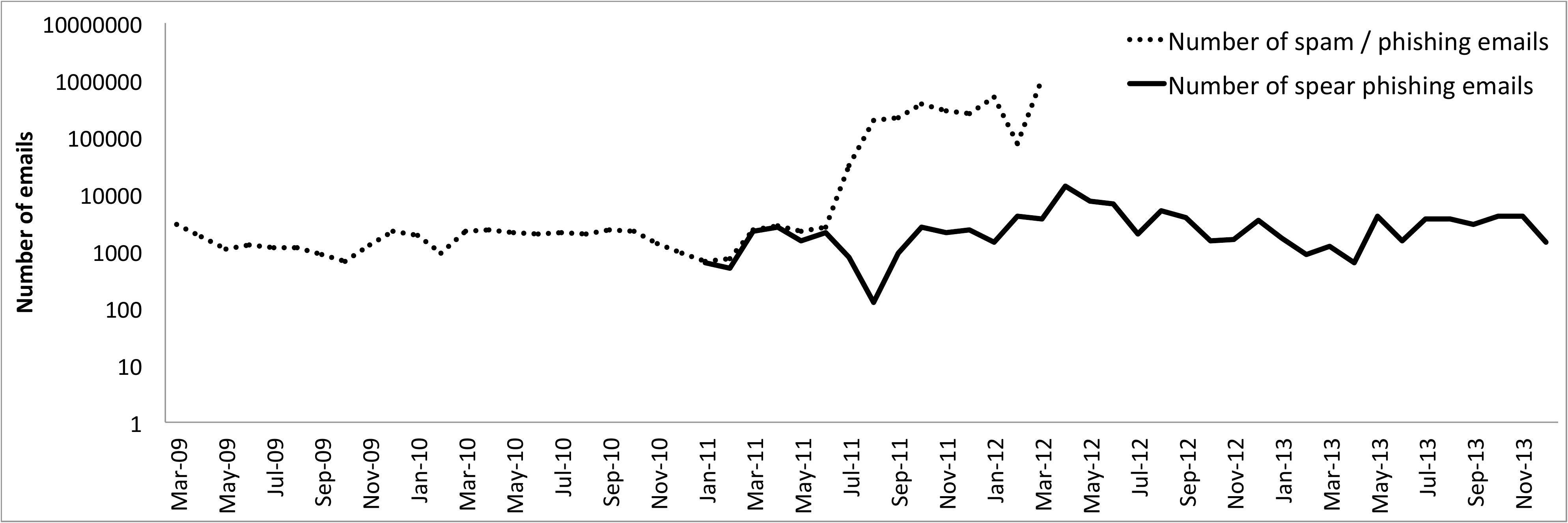}}
    \end{center}
    \caption{Timeline of the number of spear phishing and spam / phishing emails in our dataset. The X axis represents time, and the Y axis represents the number of emails on a logarithmic scale. The period of May 2011 - September 2011 saw an exponential increase in the number of spam / phishing emails in our dataset.
     }
   \label{fig:timeline}
\end{figure*}

All the emails present in our full dataset were collected over a period of slightly under 4 years, from March 2009 to December 2013. Figure~\ref{fig:timeline} presents a timeline of the ``received time" of all these emails. The spam / phishing emails were collected over a period of 3 years, from March 2009 to March 2012. The targeted spear phishing emails were also collected during a period of about 3 years, but from January 2011, to December 2013. The two datasets, thus, had data for a common time period of about 15 months, from January 2011, to March 2012. It was interesting to observe that during this period, while the number of spam and phishing emails saw a tremendous rise, the number of spear phishing emails did not vary too much. This characteristic was observed for the entire 3-year period for spear phishing emails. The number of spear phishing emails received in the beginning and end of our three year observation period saw a 238\% rise, as compared to a rise of 35,422\% percent in the number of spam / phishing emails.

In addition to the attachment information and timeline, we also looked at the ``subject" fields of all the emails present in our datasets. Table~\ref{tab:subjects} shows the top 20 most frequently occurring ``subject lines" in our datasets. Evidently, ``subjects" in both these datasets were very different in terms of context. Targeted spear phishing email subjects seemed to be very professional, talking about jobs, meetings, \emph{unclassified} information etc. Spam / phishing email subjects, however, were observed to follow a completely different genre. These emails' subjects were found to follow varied themes, out of which, three broad themes were fairly prominent: a) fake email delivery failure error messages, which lure victims into opening these emails to see which of their emails failed, and why; b) arrival of packages or couriers by famous courier delivery services -- victims tend to open such messages out of curiosity, even if they are not expecting a package; and c) personalized messages via third party websites and social networks (Hallmark E-Card, hi5 friend request, and Facebook message in this case). Most of such spam / phishing emails have generic subjects, to which most victims can relate easily, as compared to spear phishing email subjects, which would seem irrelevant to most common users.

It is important to note that these statistics are for the complete datasets we obtained from Symantec. The total number of emails present in the complete dataset was of the order of hundred thousands. However, we performed our entire analysis on a sample picked from this dataset. The analysis in the rest of the paper talks about only this sample. To make our analysis more exhaustive, we also used a sample of benign emails from the Enron email dataset for our analysis~\cite{Cohen2009}. All the three email datasets had duplicates, which we identified and removed by using a combination of 5 fields, viz. \emph{from ID}, \emph{to ID}, \emph{subject}, \emph{body}, and \emph{timestamp}. On further investigation, we found that these duplicate email messages were different instances of the same email. This happens when an email is sent to multiple recipients at the same time. A globally unique \emph{message-id} is generated for each recipient, and thus results in duplication of the message. Elimination of duplicates reduced our email sample dataset by about 50\%. Our final sample email dataset that we used for all our analysis was, therefore, a mixture of targeted attack emails, non targeted attack emails, and benign emails. We now describe this sample.

\begin{table*}[!ht]
\begin{center}
    \begin{tabular}{l|l||l|l}
    \hline
    Spear phishing subjects    & \% & Spam / phishing subjects         & \%   \\ \hline
    Job Opportunity  & 3.45 & Mail delivery failed: returning message to sender   & 10.95 \\
    Strategy Meeting    & 3.09 & Delivery Status Notification (Failure)                    & 6.71 \\
    What is Chen Guangcheng fighting for?   & 3.00 & Re:                                      & 2.59  \\
    FW: $[$2$]$ for the extension of the measures against North Korea & 1.70 & Re  & 2.56  \\
    $[$UNCLASSIFIED$]$ 2012 U.S.Army orders for weapons  & 1.27 & Become A Paid Mystery Shopper Today! Join and Shop For Free!   & 1.28  \\
 FW:$[$UNCLASSIFIED$]$2012 U.S.Army orders for weapons & 1.27 & failure notice & 1.09  \\
  $<$blank subject line$>$  & 1.17 & Delivery Status Notification (Delay)     & 1.06  \\
    FW: results of homemaking 2007 annual business plan (min quarter 1 included) & 1.02 & Returned mail: see transcript for details   & 0.95  \\
  $[$UNCLASSIFIED$]$DSO-DARPA-BAA-11-65 & 0.93   & Get a job as Paid Mystery Shopper! Shop for free and get Paid! & 0.85  \\
    Wage Data 2012  & 0.90   & Application number: AA700003125331        & 0.82  \\
   U.S.Air Force Procurement Plan 2012  & 0.90  & Your package is available for pickup & 0.78 \\
   About seconded expatriate management in overseas offices  & 0.80   & Your statement is ready for your review      & 0.75  \\
    FW:[CLASSIFIED] 2012 USA Government of the the Health Reform & 0.74   & Unpaid invoice 2913.  & 0.71  \\
    T.T COPY     & 0.62   & Track your parcel            & 0.70  \\
    USA to Provide Declassified FISA Documents & 0.58   & You have received A Hallmark E-Card!                           & 0.59  \\
    FY2011-12 Annual Merit Compensation Guidelines for Staff  & 0.55   & Your Account Opening is completed.   & 0.57  \\
    Contact List Update     & 0.45   & Delivery failure               & 0.57  \\
  DOD Technical Cooperation Program & 0.43  & Undelivered Mail Returned to Sender & 0.56  \\
  DoD Protection of Whistleblowing Spies & 0.43   & Laura would like to be your friend on hi5!                     & 0.56  \\
    FW:UK Non Paper on arrangements for the Arms Trade Treaty (ATT) Secretariat  & 0.42   & You have got a new message on Facebook!     & 0.55  \\ \hline
    \end{tabular}
\vspace{5pt}
\caption{Top 20 most frequently occurring subjects, and their corresponding percentage share in our spear phishing, and spam / phishing email datasets. Spear phishing email subjects appear to depict that these emails contain highly confidential data. Spam / phishing emails, on the other hand, are mainly themed around email delivery error messages, and courier or package receipts.}
\label{tab:subjects}
\vspace{-15pt}
\end{center}
\end{table*}

\subsection{Email Sample Dataset Description}

To focus our analysis at the organization level, we identified and extracted the most attacked organizations (excluding free email providing services like Gmail, Yahoo, Hotmail etc.) from the domain names of the victims' email addresses, and picked 14 most frequently attacked organizations. We were however, restricted to pick only those organizations, where the first names and last names were easily extractable from the email addresses. The first name and last name were required to obtain the corresponding LinkedIn profiles of these victims (this process is discussed in detail in Section~\ref{sec:linkedin_data}). This restriction, in addition to removal of duplicates, left us with a total of 4,742 targeted spear phishing emails sent to 2,434 unique victims (referred to as \emph{SPEAR} in the rest of the paper); 9,353 non targeted attack emails sent to 5,912 unique non victims (referred to as \emph{SPAM} in the rest of the paper), and 6,601 benign emails from the Enron dataset, sent to 1,240 unique Enron employees (referred to as \emph{BENIGN} in the rest of the paper). Further details of this dataset can be found in Table~\ref{tab:stats}. Table contains the number of victims, and non victims in each of the 15 companies (including Enron), and the number of emails sent to them. The victim and non victim employee sets are mutually exhaustive, and each employee in these datasets received at least one email, and had at least one LinkedIn profile. To maintain anonymity, we do not include the name of the organizations we picked; we only mention the operation sector of these companies.

\begin{table}[!ht]
\begin{center}
    \begin{tabular}{p{2.1cm}|p{0.8cm}|p{0.7cm}|p{0.8cm}|p{0.7cm}|p{1.3cm}}
    \hline
    Sector                                   & \#Victims  & \#Emails  & \#Non Victims  & \#Emails & No. of Employees \\ 
\hline
    Govt. \& Diplomatic       & 206          & 511           & 572            & 1,103   & 10,001+    \\
    Info. \& Broadcasting    & 150          & 326            & 240           & 418       & 10,001+   \\
    NGO  & 131         & 502            & 218           & 472       & 1001-5000 \\
    IT/Telecom/Defense        & 158          & 406            & 68            & 157  & 1001-5000   \\
    Pharmaceuticals                      & 120          & 216            & 589           & 862    & 10,001+    \\
    Engineering                                & 396          & 553            & 1000         & 1,625  & 10,001+  \\
    Automotive                              & 153           & 601            & 891          & 1,204  & 10,001+   \\
    Aviation/Aerospace                            & 281           & 355            & 161           & 187    & 1001-5000  \\
    Agriculture                            & 94             & 138            & 173           & 264    & 10,001+      \\
    IT \& Telecom                    & 11              & 12             & 543           & 943      & 5001-10,000  \\
    Defense                              & 388            & 651            & 123           & 147    & 10,001+    \\
    Oil \& energy                      & 201            & 212            & 680           & 1,017   & 10,001+     \\
    Finance                              & 89               & 129            & 408           & 608    & 10,001+       \\
    Chemicals                            & 56               & 130            & 248            & 346   & 10,001+     \\
    Enron                                & NA              & NA              & 1,240         & 6,601  & 10,001+        \\ \hline
    Total                                 & 2,434           &  4,742         & 7,154          & 15,954           & ~              \\ \hline
    \end{tabular}
\vspace{5pt}
\caption{Detailed description of our dataset of LinkedIn profiles and emails across 15 organizations including Enron. }
\label{tab:stats}
\vspace{-20pt}
\end{center}
\end{table}

\begin{figure*}[!ht]
     \begin{center}
        \subfigure[SPEAR subjects]{
           \label{fig:spearphish_subject}
           \includegraphics[scale=0.17]{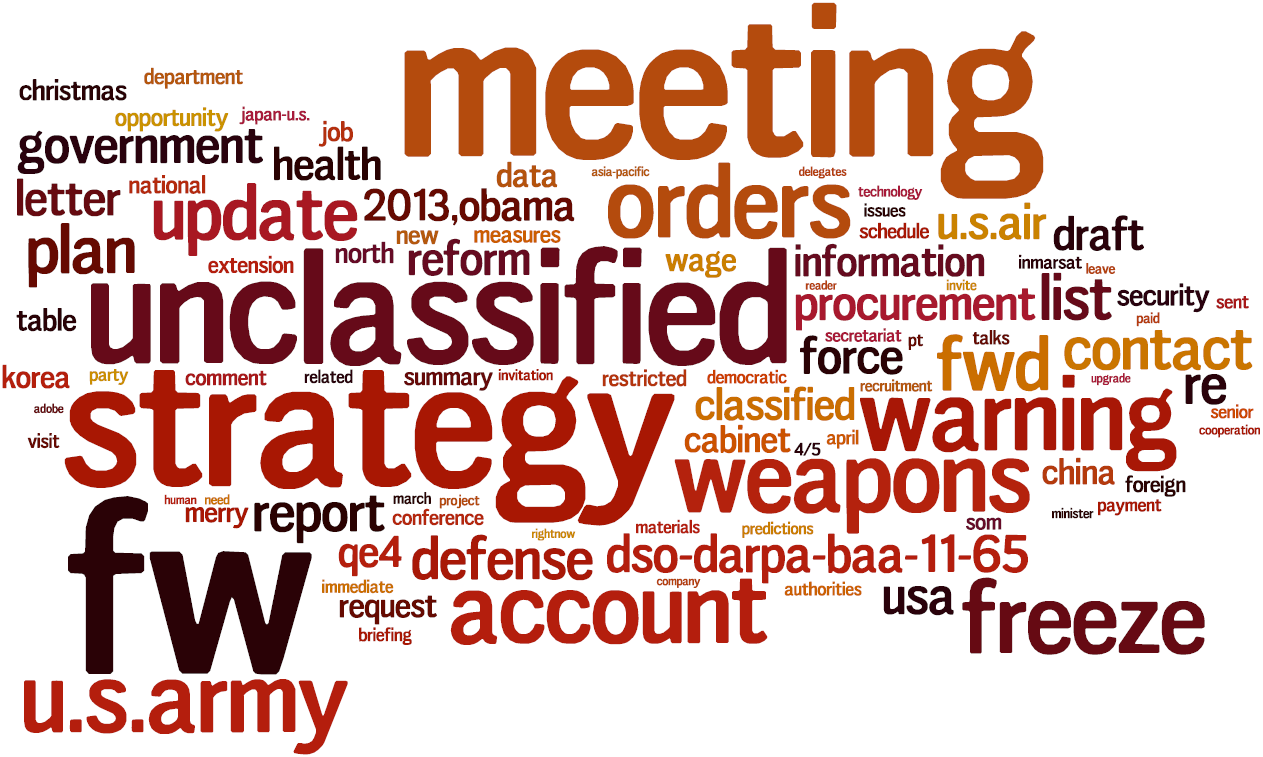}
        }
        	\subfigure[SPEAR bodies]{
            \label{fig:spearphish_body}
            \includegraphics[scale=0.16]{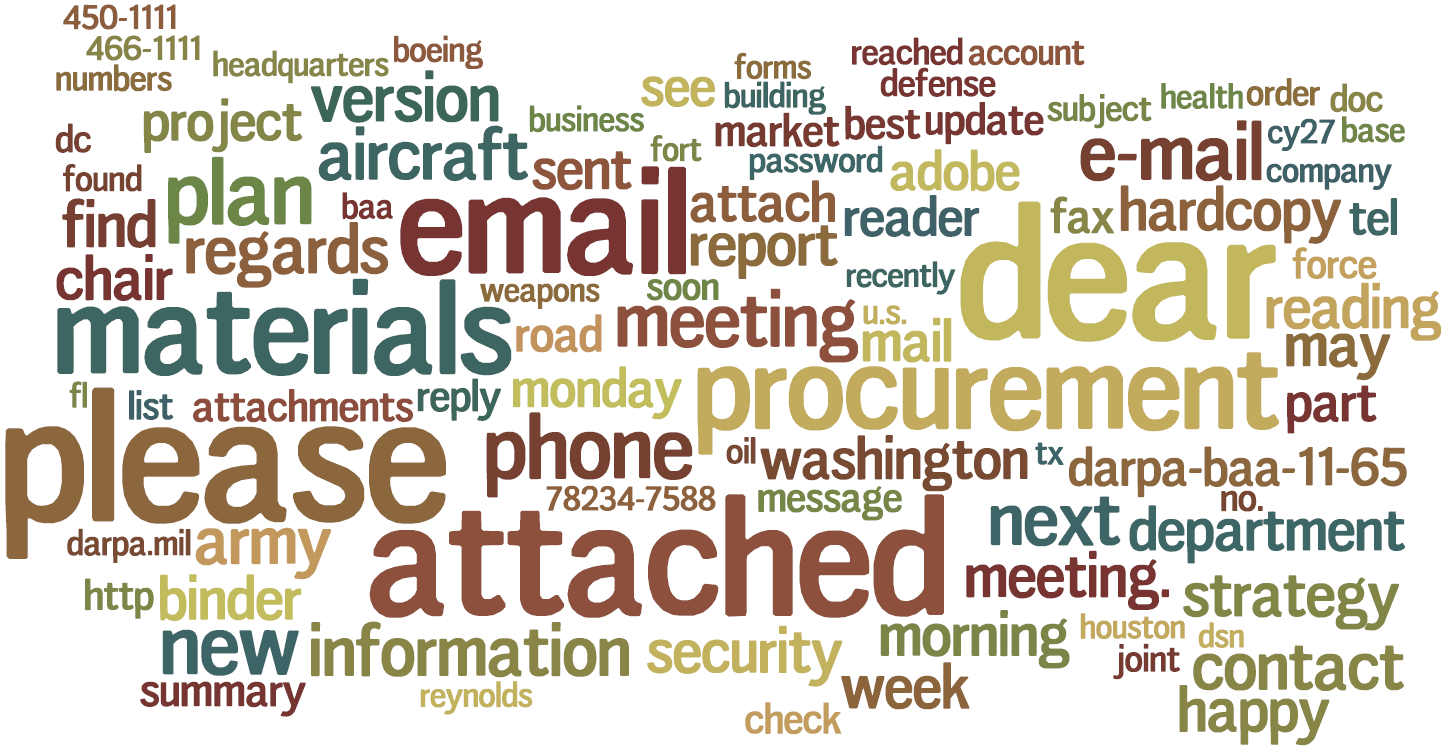}
        }
        	\subfigure[SPAM subjects]{
            \label{fig:mixed_subject}
            \includegraphics[scale=0.16]{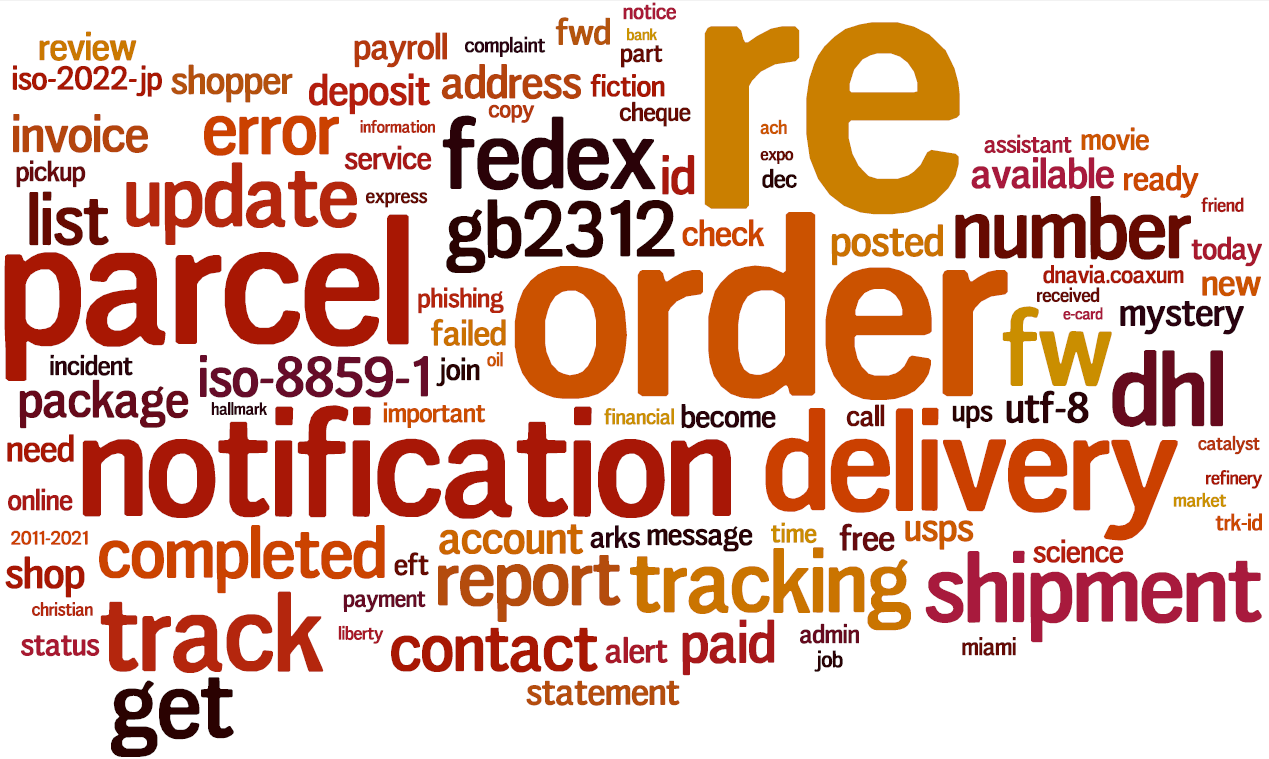}
        }\\
        	\subfigure[BENIGN subjects]{
            \label{fig:enron_subject}
            \includegraphics[scale=0.2]{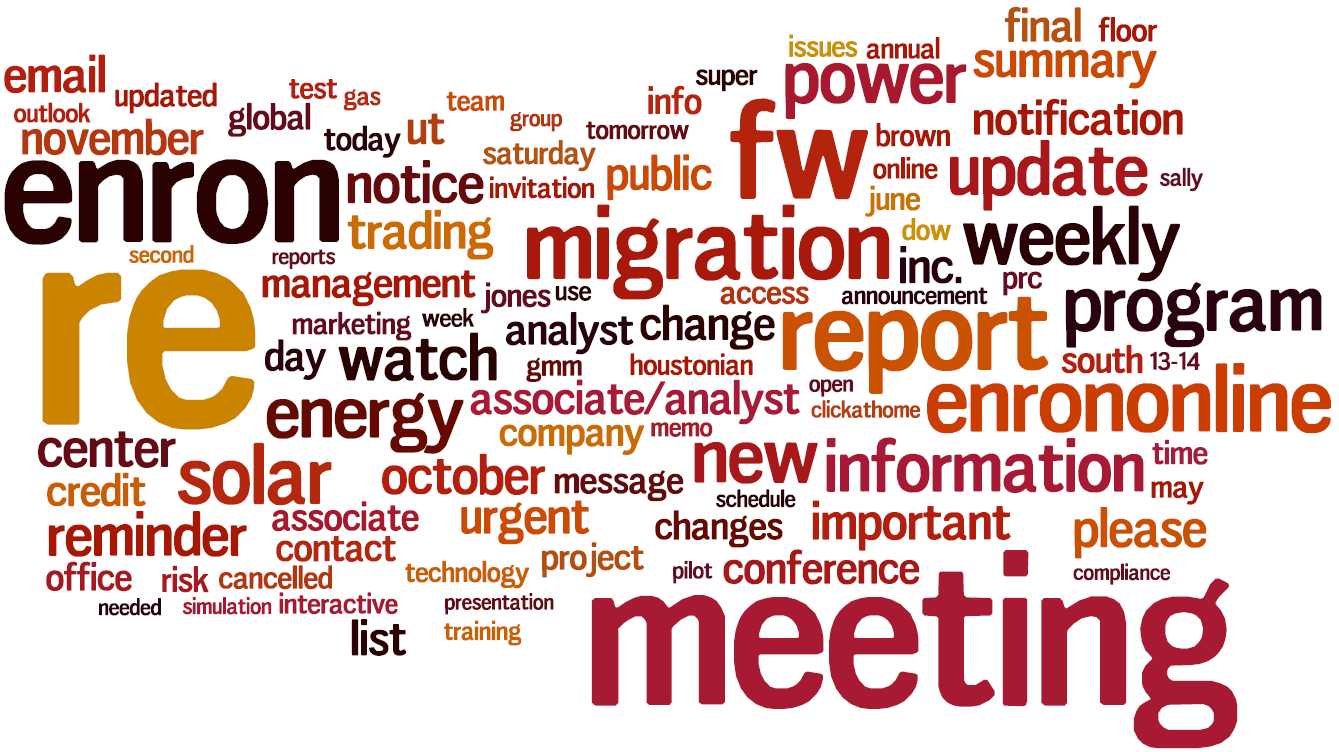}
        }
        	\subfigure[BENIGN bodies]{
            \label{fig:enron_body}
            \includegraphics[scale=0.2]{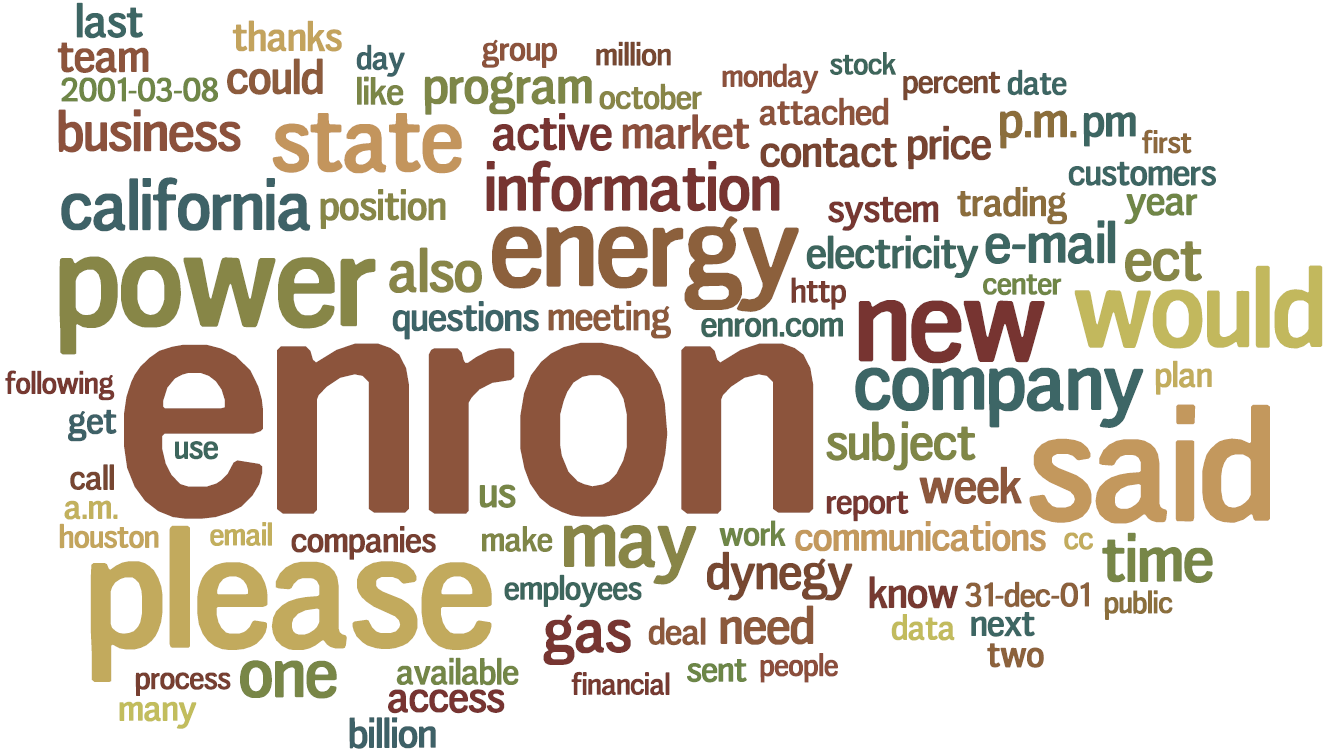}
        }
    \end{center}
    \caption{%
Tag clouds of the 100 most frequently occurring words in the subjects and bodies of our SPEAR, SPAM, and BENIGN datasets. Bodies of SPAM emails were not available in our dataset.
     }
   \label{fig:tags}
\end{figure*}

Figures~\ref{fig:spearphish_subject},~\ref{fig:mixed_subject}, and~\ref{fig:enron_subject} represent the tag clouds of the 100 most frequently occurring words in the ``subject" fields of our SPEAR, SPAM, and BENIGN datasets respectively. We noticed considerable differences between subjects from all the three datasets. While all three datasets were observed to have a lot of \emph{forwarded} emails (represented by ``fw", and ``fwd'' in the tag clouds), SPAM and BENIGN datasets were found to have much more \emph{reply} emails (signified by ``re" in the tag clouds) as compared to SPEAR emails. These characteristics of whether an email is forwarded, or a reply, have previously been used as boolean features by researchers to distinguish between phishing and benign emails~\cite{Toolan2010}. The difference in vocabularies used across the three email datasets is also notable. The SPEAR dataset (Figure~\ref{fig:spearphish_subject}) was found to be full of attention-grabbing words like \emph{strategy, unclassified, warning, weapons, defense, US Army} etc. Artifact~\ref{tab:egspear} shows an example of the attachment name, subject and body of such an email. We removed the received address and other details to maintain anonymity.

\renewcommand{\tablename}{ARTIFACT} 

\begin{table}
\begin{center}
    \begin{tabular}{|p{8.5cm}|}
\hline
    {\bf Attachment}: All information about mobile phone.rar  \\ \hline
    {\bf Subject}: RE: Issues with Phone for help    \\ \hline
    {\bf Body}: $<$name$>$,\\Thanks for your replying.I contacted my supplier,but he could not resolved it.Now I was worried, so I take the liberty of writing to you.I collect all information including sim card details,number,order record and letters in the txt file.I hope you can deal with the issues as your promised.\\Best,\\$<$name$>$\\\\-----Original Message-----\\From: Customer Care [mailto:Customer\_Care@$<$companyDomain$>$]\\Sent: 2011-12-8 0:35\\To: $<$name$>$\\Cc:\\Subject: RE: Issues with Phone for help\\\\Dear $<$name$>$,\\\\Thank you for your E-mail. I am sorry to hear of your issues. Please can you send your SIM card details or Mobile number so that we can identify your supplier who can assist you further?\\\\Thank you\\\\Kind regards,\\\\$<$name$>$\\Customer Service Executive\\\\$<$Company Name$>$,\\$<$Company Address$>$\\United Kingdom\\\\Tel: $<$telephone number$>$\\Fax : $<$Fax number$>$ \\$<$company website$>$\\\\-----Original Message-----\\From: $<$name$>$ [mailto:$<$email address$>$]\\Sent: 08 December 2011 08:27\\To: support@$<$companyDomain$>$\\Subject: Issues with Phone for help\\\\Hello,\\I purchased order for your IsatPhone Pro six months ago.Now I have trouble that it can't work normally.It often automatic shuts down.Sometimes it tells some information that i can't understand.How to do?Can you help me?\\Best,\\$<$name$>$\\\\\_\_\_\_\_\_\_\_\_\_\_\_\_\_\_\_\_\_\_\_\_\_\_\_\_\_\_\_\_\_\_\_\_\_\_\_\_\_\_\_\_\_\_\_\_\_\_\_\_\_\_\_\_\_\_\_\_\_\_\_\_\_\_\_\_\_\_\_\_\\This e-mail has been scanned for viruses by Verizon Business Internet Managed Scanning Services - powered by MessageLabs. For further information visit http://www.verizonbusiness.com/uk \\ \hline
    \end{tabular}
\vspace{5pt}
\caption{A spear phishing email from our SPEAR dataset. The email shows a seemingly genuine conversation, where the attacker sent a malicious compressed (.rar) attachment to the victim in the middle of the conversation.}
\label{tab:egspear}
\end{center}
\end{table}

SPAM emails in our dataset (Figure~\ref{fig:mixed_subject}) followed a completely different genre, dominated by words like \emph{parcel, order, delivery, tracking, notification, shipment} etc. We also found mentions of famous courier service brand names like FedEx and DHL which seem to have been used for targeting victims. Such attacks have been widely talked about in the recent past; users have also been warned about scams, and infected payloads (like spyware or malware), that accompany such emails.~\footnote{\url{http://nakedsecurity.sophos.com/2013/03/20/dhl-delivery-malware/}}~\footnote{\url{http://www.spamfighter.com/News-13360-FedEx-and-DHL-Spam-Attack-with-Greater-Ferocity.htm}} Some examples of attachment names, and subjects of such non targeted SPAM emails are shown in Artifact~\ref{tab:egspam}. BENIGN subjects comprised of diverse keywords like \emph{report, program, meeting, migration, energy}, which did not seem specific to a particular theme (Figure~\ref{fig:enron_subject}). These keywords were fairly representative of the kind of typical internal communication that may have been going on in the company.

\begin{table}[!h]
\begin{center}
    \begin{tabular}{|p{8.5cm}|}
    \hline
    {\bf Attachment}: 100A\_0.txt                                                      \\
    {\bf Subject}: DHL Express Notification for shipment  15238305825550113 \\ \hline
    {\bf Attachment}: ./attach/100\_4X\_AZ-D\_PA2\_\_FedEx=5FInvoice=5FN 56=2D141.exe     \\
    {\bf Subject}: FEDEX Shipment Status NR-6804 \\ \hline
    \end{tabular}
\vspace{5pt}
\caption{Examples of \emph{subject} and \emph{attachment} names of two spam emails from our SPAM dataset. The \emph{body} field of the emails was not available in this dataset.}
\label{tab:egspam}
\end{center}
\end{table}

\renewcommand{\tablename}{TABLE}

We also compared the body content of SPEAR and BENIGN emails. Figures~\ref{fig:spearphish_body} and~\ref{fig:enron_body} represent the tag clouds of the 100 most frequently occurring words in the body fields of the SPEAR and BENIGN datasets respectively. Contrary to our observations from the subject content in the SPEAR dataset (Figure~\ref{fig:spearphish_subject}), the body content of the SPEAR emails (Figure~\ref{fig:spearphish_body}) did not look very attention-grabbing or themed. SPEAR bodies contained words like \emph{attached, please, email, dear, materials, phone} etc., which commonly occur in routine email communications too. The BENIGN body content did not contain anything peculiar or alarming either (Figure~\ref{fig:enron_body}). Since Symantec's email dataset of spear phishing, spam and phishing emails isn't publicly available, we believe that this characterization of our dataset can be useful for researchers to get a better idea of state-of-the-art, real world malicious email data which circulates in the corporate environment. 

%
%

\subsection{LinkedIn profile dataset} \label{sec:linkedin_data}

Our second dataset consisted of LinkedIn profiles of the receivers of all the emails present in our email dataset. In fact, we restricted our email dataset to only those emails which were sent to employees having at least one LinkedIn profile. This was done to have a complete dataset in terms of the availability of social and stylometric features. There were two major challenges with data collection from LinkedIn; a) Strict input requirements, and b) Rate limited API.

Firstly, to fetch the profiles of LinkedIn users who are outside a user's network (3$^{rd}$ degree connections and beyond), the LinkedIn People Search API requires first name, last name, and company name as a mandatory input.~\footnote{\url{developer.linkedin.com/documents/people-search-api}} Understandably, none of the users we were looking for, were in our network, and thus, as specified in the previous subsection, we were restricted to pick up emails of only those companies which followed the format \emph{firstName}.\emph{lastName}$@$\emph{companyDomain} or \emph{firstName}\_\emph{lastName}$@$\emph{companyDomain}. Restricting our dataset to such email addresses was the only way we could satisfy the API's input requirements.

\begin{figure}[!h]
     \begin{center}
\includegraphics[scale=0.36]{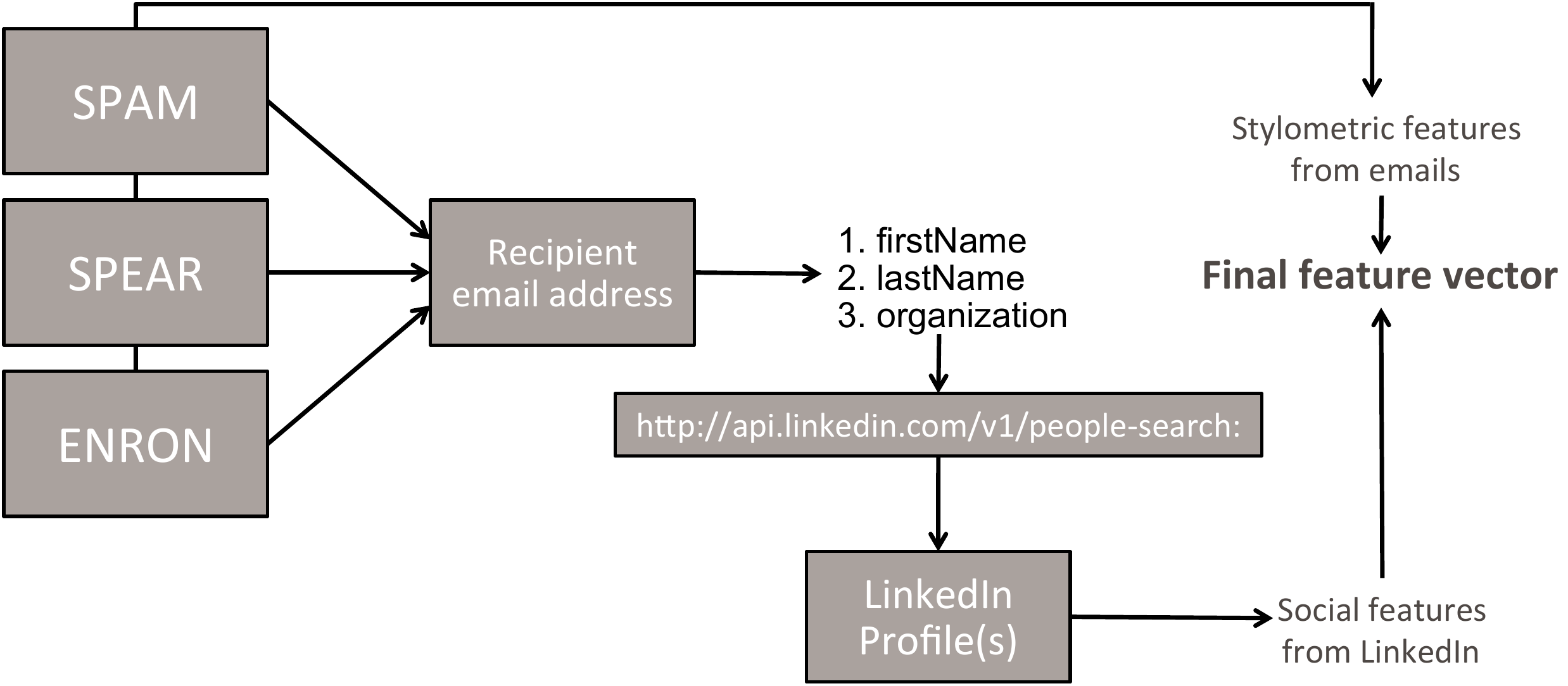}
    \end{center}
    \caption{%
Flow diagram describing the data collection process we used to collect LinkedIn data, and create our final feature vector containing stylometric features from emails, and social features from LinkedIn profiles.
     }
   \label{fig:arch}
\end{figure}

Secondly, the rate limit of the people search API posed a major hindrance. Towards the end of 2013, LinkedIn imposed a tight limit of 250 calls per day, per application, on the people search API for existing developers, and completely restricted access for new applications and developers, under their Vetted API access program.~\footnote{\url{https://developer.linkedin.com/blog/vetted-api-access}} We 
were able to get access to the Vetted API for two of our applications. Although the new rate limit allowed 100,000 API calls per day, per application, this was still restricted to 100 calls per user, per day, per application. We then created multiple LinkedIn user accounts to make calls to the API. Even with multiple applications, and user accounts, this data collection process took about 4 months. This happened because a lot of our search queries to the API returned no results. On average, we were able to find a LinkedIn profile for only 1 in 10 users in our dataset. This resulted in about 90\% of the API calls returning no results, and hence getting wasted. Eventually, we were able to collect a total of 2,434 LinkedIn profiles of victims, 5,914 LinkedIn profiles of non victims, across the 14 organizations; and 1,240 LinkedIn profiles of employees from Enron (Table~\ref{tab:stats}). To obtain these profiles for the 9,588 employees (2,434 victims, 5,914 non victims, and 1,240 Enron employees), the number of API calls we had to make was approximately 100,000 (approx. 10 times the number of profiles obtained). Figure~\ref{fig:arch} shows the flow diagram describing our data collection process.

Our first choices for extracting \emph{social} features about employees were Facebook, and Twitter. However, we found that identifying an individual on Facebook and Twitter using only the first name, last name, and employer company was a hard task. Unlike LinkedIn, the Facebook and Twitter APIs do not provide endpoints to search people using the work company name. This left us with the option to search for people using first name, and last name only. However, searching for people using only these two fields returned too many results on both Facebook and Twitter, and we had no way to identify the correct user that we were looking for. We then visited the profile pages of some users returned by the API results manually, and discovered that the \emph{work} field for most users on Facebook was private. On Twitter profiles, there did not exist a \emph{work} field at all. It thus became very hard to be able to find Facebook or Twitter profiles using the email addresses in our dataset. 


\section{Analysis and results} \label{sec:ar}

To distinguish spear phishing emails from non spear phishing emails using \emph{social} features of the receivers, we used four machine learning algorithms, and a total of 27 features; 18 stylometric, and 9 \emph{social}. The entire analysis and classification tasks were performed using the Weka data mining software~\cite{Hall2009}. We applied 10-fold cross validation to validate our classification results. We now describe our feature sets, analysis, and results of the classification.

\subsection{Feature set description} \label{sec:fsd}

We extracted a set of 18 stylometric features from each email in our email dataset, and a set of 9 \emph{social} features from each LinkedIn profile present in our LinkedIn profile dataset, features described in Table~\ref{tab:feats}. Features extracted from our email dataset are further categorized into three categories, viz. \emph{subject} features, \emph{attachment} features, and \emph{body} features. It is important to note that we did not have all the three types of these aforementioned features available for all our datasets. While the SPAM dataset did not have \emph{body} features, the BENIGN dataset did not have the \emph{attachment} features. Features marked with $^*$ (in Table~\ref{tab:feats}) have been previously used by researchers to classify spam and phishing emails~\cite{Toolan2010}. The \emph{richness} feature is calculated as the ratio of the number of words to the number of characters present in the text content under consideration. We calculate richness value for the email \emph{subject}, email \emph{body}, and LinkedIn profile \emph{summary}. The \emph{Body\_hasAttach} features is a boolean variable which is set to true, if the body of the email contain the word ``attached'' or ``attachment'', indicating that an attachment is enclosed with the email. This feature helped us to capture the presence of attachments for the BENIGN dataset, which did not have attachment information. The \emph{Body\_numFunctionWords} feature captures the total number of occurrences of function words present in the email body, from a list of function words which includes the words: ~\emph{account, access, bank, credit, click, identity, inconvenience, information, limited, log, minutes, password, recently, risk, social, security, service,} and \emph{suspended}. These features have been previously used by Chandrasekaran~\cite{Chandrasekaran2006}.

\begin{table}[!ht]
\begin{center}
    \begin{tabular}{l|l|l}
\hline
    Feature                                & Data Type                         & Source                            \\ \hline
    Subject\_IsReply$^*$             & Boolean                            & Email                              \\
    Subject\_hasBank$^*$            & Boolean                            & Email                               \\
    Subject\_numWords$^*$         & Numeric                            & Email                              \\
    Subject\_numChars$^*$          & Numeric                            & Email                               \\
    Subject\_richness$^*$             & Decimal (0-1)                     & Email                               \\
    Subject\_isForwarded$^*$        & Boolean                             & Email                               \\
    Subject\_hasVerify$^*$            & Boolean                             & Email                               \\ \hline
    Length of attachment name       & Numeric                             & Email                                \\
    Attachment size (bytes)           & Numeric                            & Email                                \\ \hline
    Body\_numUniqueWords$^*$    & Numeric                             & Email                               \\
    Body\_numNewlines                 & Numeric                             & Email                                 \\
    Body\_numWords$^*$             & Numeric                             & Email                                 \\
    Body\_numChars$^*$              & Numeric                             & Email                                \\
    Body\_richness$^*$                 & Decimal (0-1)                      & Email                                \\
    Body\_hasAttach                     & Boolean                             & Email                                \\
    Body\_numFunctionWords$^*$   & Numeric                            & Email                               \\
    Body\_verifyYourAccount$^*$    & Boolean                             & Email                               \\
    Body\_hasSuspension$^*$         & Boolean                              & Email                               \\ \hline 
    Location                                 & Text (country)                       & LinkedIn                         \\
    numConnections                      & Numeric (0-500)                    & LinkedIn                          \\
    SummaryLength                      & Numeric                               & LinkedIn                          \\
    SummaryNumChars                 & Numeric                               & LinkedIn                           \\
    SummaryUniqueWords              & Numeric                              & LinkedIn                            \\
    SummaryNumWords                 & Numeric                               & LinkedIn                           \\
    SummaryRichness                    & Decimal (0-1)                         & LinkedIn                          \\
    jobLevel                                 & Numeric (0-7)                         & LinkedIn                         \\
    jobType                                 & Numeric (0-9)                         & LinkedIn                          \\
 \hline
    \end{tabular}
\vspace{5pt}
\caption{List of features used in our analysis. We used a combination of stylometric features in addition to normal features. Features marked with $^*$ have been previously used for detecting spam and phishing emails.}
\label{tab:feats}

\end{center}
\end{table}

The \emph{social} features we extracted from the LinkedIn profiles, captured three distinct types of information about an employee, viz. location, connectivity, and profession. The \emph{Location} was a text field containing the state / country level location of an employee, as specified by her on her LinkedIn profile. We extracted and used the country for our analysis. The \emph{numConnections} was a numeric field, and captured the number of connections that a user has on LinkedIn. If the number of connections for a user is more than 500, the value returned is ``500+" instead of the actual number of connections. These features captured the location and connectivity respectively. In addition to these two, we extracted 5 features from the \emph{Summary} field, and 2 features from the \emph{headline} field returned by LinkedIn's People Search API. The \emph{Summary} field is a long, free-text field comprising of a summary about a user, as specified by her, and is optional. The features we extracted from this field were similar to the ones we extracted from the subject and body fields in our email dataset. These features were, the \emph{summary length, number of characters, number of unique words, total number of words}, and \emph{richness}. We introduced two new features, \emph{job\_level} and \emph{job\_type}, which are numeric values ranging from 0 to 7, and 0 to 9 respectively, describing the position and area of work of an individual. We looked for presence of certain level and designation specific keywords in the ``headline'' field of a user, as returned by the LinkedIn API. The job levels and job types, and their numeric equivalents are as follows:

\begin{itemize}
\item Job\_level; maximum of the following:

1 - Support\\
2 - Intern\\
3 - Temporary\\
4 - IC\\
5 - Manager\\
6 - Director\\
7 - Executive\\
0 - Other; if none of the above are found.

\item Job\_type; minimum of the following:

1 - Engineering\\
2 - Research\\
3 - QA\\
4 - Information Technology\\
5 - Operations\\
6 - Human Resources\\
7 - Legal\\
8 - Finance\\
9 - Sales / Marketing\\
0 - Other; if none of the above are found.
\end{itemize}

To see if information extracted about a victim from online social media helps in identifying a spear phishing email sent to her, we performed classification using a) \emph{email} features~\footnote{We further split email features into \emph{subject}, \emph{body}, and \emph{attachment} features for analysis, wherever available.}; b) \emph{social} features, and c) using a combination of these features. We compared these three accuracy scores across a combination of datasets viz. SPEAR versus SPAM emails from Symantec's email scanning service, SPEAR versus benign emails from BENIGN dataset, and SPEAR versus a mixture of emails from BENIGN, and SPAM from the Symantec dataset. As mentioned earlier, not all \emph{email} features mentioned in Table~\ref{tab:feats} were available for all the three email datasets. The BENIGN dataset did not have attachment related features, and the \emph{body} field was missing in the SPAM email dataset. We thus used only those features for classification, which were available in both the targeted, and non targeted emails.

\subsection{SPEAR versus SPAM emails from Symantec} \label{sec:sp_vs_spam}

Table~\ref{tab:sp_spam} presents the results of our first analysis where we subjected SPEAR and SPAM emails from Symantec, to four machine learning classification algorithms, viz. Random Forest~\cite{Breiman2001}, J48 Decision Tree~\cite{Quinlan1993}, Naive Bayesian~\cite{John1995}, and Decision Table~\cite{Kohavi1995}. Feature vectors for this analysis were prepared from 4,742 SPEAR emails, and 9,353 SPAM emails, combined with \emph{social} features extracted from the LinkedIn profiles of receivers of these emails. Using a combination of all \emph{email} and \emph{social} features, we were able to achieve a maximum accuracy of 96.47\% using the Random Forest classifier for classifying SPEAR and SPAM emails. However, it was interesting to notice that two out of the four classifiers performed better \emph{without} the social features. Although the Decision Table classifier seemed to perform equally well with, and without the social features, it performed much better using only \emph{email} features, as compared to only \emph{social} features.~\footnote{This happened because the Decision Table classifier terminates search after scanning for a certain (fixed) number of non-improving nodes / features.} In fact, the Decision Table classifier achieved the maximum accuracy using \emph{attachment} features, which highlights that the attachments associated with SPEAR and SPAM emails were also substantially different in terms of name and size. We achieved an overall maximum accuracy of 98.28\% using the Random Forest classifier trained on only email features. This behavior revealed that the public information available on the LinkedIn profile of an employee in our dataset, does not help in determining whether she will be targeted for a spear phishing attack or not.

\begin{table}[!h]
\begin{center}
    \begin{tabular}{l|l|p{0.7cm}|p{1cm}|p{0.9cm}|p{0.7cm}}
    \hline
    Feature set & Classifier   & Random Forest & J48 Decision Tree & Naive Bayesian & Decision Table \\ \hline
    Subject (7)              & Accuracy (\%) & 83.91         & 83.10             & 58.87    & 82.04      \\
    ~                        & FP rate     & 0.208         & 0.227             & 0.371     & 0.227       \\ \hline
    Attachment (2)           & Accuracy (\%) & 97.86         & 96.69             & 69.15   & {\bf 95.05}       \\
    ~                        & FP rate     & 0.035         & 0.046             & 0.218   & 0.056       \\ \hline
    All email (9)           & Accuracy (\%) & {\bf 98.28}   & {\bf 97.32}     & 68.69    & {\bf 95.05}      \\
    ~                        & FP rate           & 0.024          & 0.035              & 0.221    & 0.056      \\ \hline
    Social (9)            & Accuracy (\%) & 81.73         & 76.63             & 65.85    & 70.90      \\
    ~                        & FP rate     & 0.229         & 0.356             & 0.445  & 0.41        \\ \hline
    Email +     & Accuracy (\%) & 96.47         & 95.90             & {\bf 69.35}   & {\bf 95.05}    \\
    Social (18)                        & FP rate     & 0.052         & 0.054             & 0.232     & 0.056      \\ \hline
    \end{tabular}
\vspace{5pt}
\caption{Accuracy and weighed false positive rates for SPEAR versus SPAM emails. Social features reduce the accuracy when combined with email features.}
\label{tab:sp_spam}
\end{center}
\end{table}

To get a better understanding of the results, we looked at the information gain associated with each feature using the InfoGainAttributeEval Attribute Evaluator package.~\footnote{\url{http://weka.sourceforge.net/doc.dev/weka/attributeSelection/InfoGainAttributeEval.html}} This package calculates the ~\emph{information gain}~\footnote{This value ranges between 0 and 1, where a higher value represents a more discriminating feature.} associated with each feature, and ranks the features in descending order of the information gain value. The ranking revealed that the attachment related features were the most distinguishing features between SPEAR and SPAM emails. This phenomenon was also highlighted by the Decision Table classifier (Table~\ref{tab:sp_spam}). The attachment size was the most distinguishing feature with an information gain score of 0.631, followed by length of attachment name, with an information gain score of 0.485. As evident from Table~\ref{tab:rankedfeats1}, attachment sizes associated with SPAM emails have very high standard deviation values, even though the average attachment sizes of SPAM and SPEAR emails are fairly similar. It is also evident that attachments associated with SPAM emails tend to have longer names; on average, twice in size as compared to attachments associated with SPEAR emails. Among subject features, we found no major difference in the length (number of characters, and number of words) of the subject fields across the two email datasets.

\begin{table}[!h]
\begin{center}
    \begin{tabular}{l|l|c|c|c|c}
    \hline
    \multirow{2}{*}{Feature}    & \multirow{2}{*}{Info. Gain} & \multicolumn{2}{c|}{SPEAR} & \multicolumn{2}{c}{SPAM} \\ \cline{3-6}
    ~                         & ~          & Mean           & Std Dev. & Mean & Std. Dev. \\ \hline
    Attachment size (Kb)   & 0.6312   & 285              & 531        & 262    & 1,419         \\
    Len. attachment name & 0.4859   & 25.48              & 16.03        & 51.08    & 23.29         \\
    Subject\_richness         & 0.2787   & 0.159              & 0.05        & 0.177    & 0.099         \\
    Subject\_numChars         & 0.1650   & 29.61              & 17.77        & 31.82    & 23.85         \\
    Location                  & 0.0728   & -              & -        & -    & -         \\
    Subject\_numWords         & 0.0645   & 4.74              & 3.28        & 4.59    & 3.97         \\
    numConnections            & 0.0219          & 158.68              & 164.31        & 183.82    & 171.45         \\
    Subject\_isForwarded      & 0.0219   & -              & -        & -    & -        \\ 
    Subject\_isReply            & 0.0154    & -              & -        & -    & -         \\
    SummaryRichness          & 0.0060    & 0.045            & 0.074        & 0.053    & 0.078      \\ \hline
    \end{tabular}
\vspace{5pt}
\caption{Information gain, mean and standard deviation of the 10 most informative features from SPEAR and SPAM emails.}
\label{tab:rankedfeats1}
\end{center}
\end{table}

\begin{figure}[!h]
     \begin{center}
\includegraphics[scale=0.42]{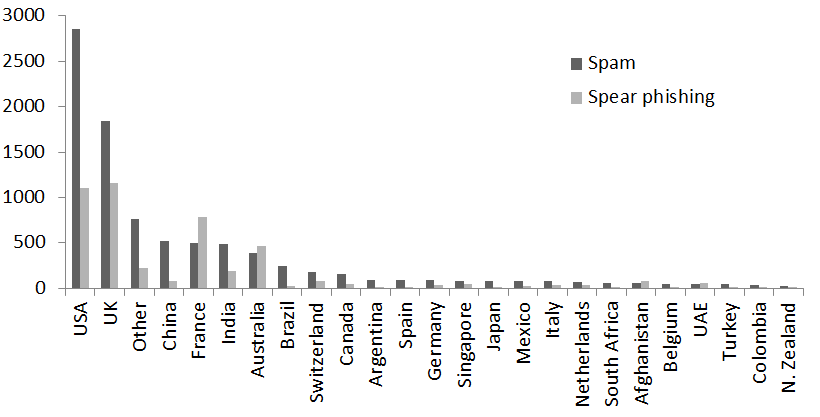}
    \end{center}
    \caption{%
Number of SPEAR and SPAM emails received by employees in the top 25 locations extracted from their LinkedIn profiles. Employees working in France, Australia, and Afghanistan received more SPEAR emails than SPAM emails.
     }
   \label{fig:countries_spam_sp}
\end{figure}

It was interesting to see that apart from the Location, number of LinkedIn connections, and SummaryRichness, none of the other social features were ranked amongst the top 10 informative features. Figure~\ref{fig:countries_spam_sp} shows the top 25 \emph{Locations} extracted from the LinkedIn profiles of employees of the 14 companies who received SPAM and SPEAR emails. We found a fairly high correlation of 0.88 between the number of SPAM and SPEAR emails received at these locations, depicting that there is not much difference between the number of SPAM and SPEAR emails received by most locations. This falls in line with the low information gain associated with this feature. Among the top 25, only 3 locations viz. France, Australia, and Afghanistan received more SPEAR emails than SPAM emails.

\begin{figure}[!h]
     \begin{center}
\includegraphics[scale=0.33]{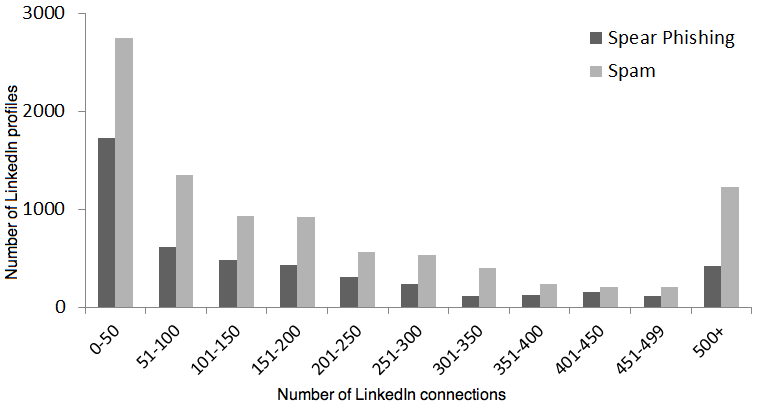}
    \end{center}
    \caption{%
Number of LinkedIn connections of the recipients of SPEAR and SPAM emails. The number of connections are plotted on the X axis, and the number of employee profiles are plotted on the Y axis. Maximum number of employee profiles had less than 50 LinkedIn connections.
     }
   \label{fig:linkedin1}
\end{figure}

The number of LinkedIn connections of the recipients of SPEAR and SPAM emails in our dataset are presented in Figure~\ref{fig:linkedin1}. There wasn't much difference between the number of LinkedIn connections of recipients of SPEAR emails, and the number of LinkedIn connections of recipients of SPAM emails. We grouped the number of LinkedIn connections into 11 buckets as represented by the X axis in Figure~\ref{fig:linkedin1}, and found a strong correlation value of 0.97 across the two classes (SPEAR and SPAM). This confirmed that the number of LinkedIn connections did not vary much between recipients of SPEAR and SPAM emails, and thus, is not an informative feature for distinguishing between SPEAR and SPAM emails.

\subsection{SPEAR emails versus BENIGN emails}

Similar to the analysis performed in Section~\ref{sec:sp_vs_spam}, we applied machine learning algorithms on a different dataset containing SPEAR emails, and BENIGN emails. This dataset contained 4,742 SPEAR emails, and 6,601 benign emails from BENIGN. Since BENIGN mostly contains internal email communication between Enron's employees, we believe that it is safe to assume that none of these emails would be targeted spear phishing emails, and can be marked as benign. Similar to our observations in Section~\ref{sec:sp_vs_spam}, we found that, in this case too, only \emph{email} features performed slightly better than a combination of \emph{email} and \emph{social} features, at distinguishing spear phishing emails from non spear phishing emails. We were able to achieve a maximum accuracy of 97.04\% using the Random Forest classifier trained on a set of 25 features; 16 \emph{email}, and 9 \emph{social} features. However, the overall maximum accuracy that we were able to achieve for this dataset was 97.39\%, using only \emph{email} features. Table~\ref{tab:sp_enron} shows the results of our analysis in detail. Three out of the four classifiers performed best with \emph{email} features; two classifiers performed best using a combination of \emph{subject} and \emph{body} features, while one classifier performed best using only \emph{body} features. The Naive Bayes classifier worked best using \emph{social} features.

\begin{table}[!h]
\begin{center}
    \begin{tabular}{l|l|p{0.8cm}|p{1cm}|p{0.9cm}|p{0.7cm}}
    \hline
    Feature set & Classifier & Random Forest & J48 Decision Tree & Naive Bayesian & Decision Table \\ \hline
    Subject (7)              & Accuracy (\%) & 81.19         & 81.11             & 61.75  & 79.55        \\
    ~                        & FP rate     & 0.210         & 0.217             & 0.489   & 0.228       \\ \hline
    Body (9)           & Accuracy (\%) & 97.17         & 95.62             & 53.81    & {\bf 90.85}      \\
    ~                        & FP rate     & 0.031         & 0.048             & 0.338    & 0.082      \\ \hline
    All email (16)           & Accuracy (\%) & {\bf 97.39}         & {\bf 95.84}  & 54.14   & 89.80       \\
    ~                        & FP rate     & 0.029          & 0.044              & 0.334   & 0.090       \\ \hline
    Social (9)            & Accuracy (\%) & 94.48         & 91.79             & {\bf 69.76}  & 83.80        \\
    ~                        & FP rate     & 0.067         & 0.103             & 0.278    & 0.198      \\ \hline
    Email +     & Accuracy (\%) & 97.04         & 95.28             & 57.27     & 89.80     \\
    Social (25)       & FP rate     & 0.032         & 0.052             & 0.316    & 0.090       \\ \hline
    \end{tabular}
\vspace{5pt}
\caption{Accuracy and weighed false positive rates for SPEAR emails versus BENIGN emails. Similar to SPEAR versus SPAM, social features decrease the accuracy when combined with email features in this case too.}
\label{tab:sp_enron}
\end{center}
\end{table}

Table~\ref{tab:rankedfeats2} presents the 10 most informative features, along with their information gain, mean and standard deviation values from the SPEAR and BENIGN datasets. The \emph{body} features were found to be the most informative in this analysis, with only 2 \emph{social} features among the top 10. Emails in the BENIGN dataset were found to be much longer than SPEAR emails in our Symantec dataset in terms of number of words, and number of characters in their ``body". The ``subject" lengths, however, were found to be very similar across SPEAR and BENIGN.

\begin{table}[!h]
\begin{center}
    \begin{tabular}{p{2.5cm}|p{0.6cm}|c|c|c|c}
    \hline
    \multirow{2}{*}{Feature}    & Info. & \multicolumn{2}{c|}{SPEAR} & \multicolumn{2}{c}{BENIGN} \\ \cline{3-6}
    ~                                &  Gain              & Mean         & Std Dev.             & Mean        & Std. Dev. \\ \hline
    Body\_richness    & 0.6506     & 0.134             &  0.085           & 0.185              & 0.027          \\
    Body\_numChars     & 0.5816     & 313.60       & 650.48           & 1735.5              & 8692.6     \\
    Body\_numWords          &  0.4954    & 53.12   & 107.53           & 312.81              & 1572.1    \\
    Body\_numUniqueWords & 0.4766     & 38.08     & 49.70       & 149.93              & 416.40      \\
    Location                       & 0.3013     & -                    & -               & -                & -              \\
    Body\_numNewlines    & 0.2660      & 11.29     & 32.70           & 43.58               & 215.77      \\
    Subject\_richness   & 0.2230     & 0.159           & 0.051           & 0.174              & 0.056         \\
    numConnections      & 0.1537     & 158.68      & 164.31         & 259.89             & 167.14       \\
    Subj\_numChars    & 0.1286     & 29.61               & 17.77            & 28.54             & 15.23  \\ 
    Body\_numFunctionWords& 0.0673     & 0.375     & 1.034          & 1.536             & 5.773      \\
    \hline
    \end{tabular}
\vspace{5pt}
\caption{Information gain, mean and standard deviation of the 10 most informative features from SPEAR and BENIGN emails. The \emph{body} features performed best at distinguishing SPEAR emails from BENIGN emails.}
\label{tab:rankedfeats2}
\end{center}
\end{table}

The Random Forest classifier was also able to achieve an accuracy rate of 94.48\% using only \emph{social} features; signifying that there exist distinct differences between the LinkedIn profiles of Enron employees, and the LinkedIn profiles of the employees of the 14 companies in our dataset. The \emph{location} attribute was found to be the most distinguishing feature among the \emph{social} features. This was understandable since most of the Enron employees were found to be based in the US (as Enron was an American services company). However, we also found a considerable difference in the average number of LinkedIn connections of Enron employees, and employees of the 14 organizations from our dataset (mean values for \emph{numConnections} feature in Table~\ref{tab:rankedfeats2}).

\subsection{SPEAR versus a mixture of BENIGN and SPAM}

While analyzing SPEAR with SPAM, and BENIGN emails separately, we found similar results where \emph{social} features were not found to be very useful in both the cases. So we decided to use a mixture of SPAM and BENIGN emails against SPEAR emails, and perform the classification tasks again. We found that in this case, two out of the four classifiers performed better with a combination of email and social features, while two classifiers performed better with only \emph{email} features. However, the overall maximum accuracy was achieved using a combination of \emph{email} and \emph{social} features (89.86\% using Random Forest classifier). This result is in contradiction with our analysis of SPEAR versus SPAM, and SPEAR versus BENIGN separately, where \emph{email} features always performed better independently, than a combination of \emph{email} and \emph{social} features. Our overall maximum accuracy, however, dropped to 89.86\% (from 98.28\% in SPEAR versus SPAM email classification) because of the absence of \emph{attachment} features in this dataset. Although the \emph{attachment} features were available in the SPAM dataset, their unavailability in BENIGN forced us to remove this feature for the current classification task. Eventually, merging the SPAM email dataset with BENIGN reduced our email dataset to only 7 features, all based on the email ``subject". Table~\ref{tab:sp_spamenron} presents the detailed results from this analysis.

\begin{table}[!h]
\begin{center}
    \begin{tabular}{l|l|p{0.8cm}|p{1cm}|p{0.9cm}|p{0.7cm}}
    \hline
    Feature set & Classifier      & Random Forest & J48 Decision Tree & Naive Bayesian & Decision Table \\ \hline
    Subject (7)              & Accuracy (\%) & 86.48         & 86.35             & {\bf 77.99}    &  {\bf 85.46}      \\
    ~                        & FP rate     & 0.333        & 0.352            & 0.681   &    0.341    \\ \hline
    Social (9)            & Accuracy (\%) & 88.04         & 84.69             & 74.46  & 80.61       \\
    ~                        & FP rate     & 0.241         & 0.371             & 0.454    & 0.432      \\ \hline
    Email +     & Accuracy (\%) & {\bf 89.86}         & {\bf 88.38}             & 73.97  & 84.14        \\
    Social (16)    & FP rate     & 0.202         & 0.248             & 0.381   & 0.250        \\ \hline
    \end{tabular}
\vspace{5pt}

\caption{Accuracy and weighed false positive rates for SPEAR emails versus mix of SPAM emails and BENIGN emails. Unlike SPEAR versus SPAM, or SPEAR versus BENIGN, \emph{social} features increased the accuracy when combined with email features in this case.}
\label{tab:sp_spamenron}
\end{center}
\end{table}

As mentioned earlier, combining the SPAM email dataset with BENIGN largely reduced our \emph{email} feature set. We were left with 7 out of a total of 18 email features described in Table~\ref{tab:feats}. Understandably, due to this depleted \emph{email} feature set, we found that the email features did not perform as good as \emph{social} features in this classification task. Despite being fewer in number, the \emph{subject} features, viz. \emph{Subject\_richness} and \emph{Subject\_numChars} were found to be two of the most informative features (Table~\ref{tab:rankedfeats3}). However, the information gain value associated with both these features was fairly low. This shows that even being the best features, the \emph{Subject\_richness} and \emph{Subject\_numChars} were not highly distinctive features amongst spear phishing, and non spear phishing emails. Similar mean and standard deviation values for both these features in Table~\ref{tab:rankedfeats3} confirm these outcomes.

\begin{table}[!h]
\begin{center}
    \begin{tabular}{l|l|c|c|c|c}
    \hline
    \multirow{2}{*}{Feature}    & \multirow{2}{*}{Info. Gain} & \multicolumn{2}{c|}{SPEAR} & \multicolumn{2}{c}{SPAM + BENIGN} \\ \cline{3-6}
    ~                            & ~            & Mean           & Std Dev. & Mean & Std. Dev. \\ \hline
    Subject\_richness       & 0.1829      & 0.159              & 0.051        & 0.176    & 0.084         \\
    Subject\_numChars     & 0.1050     & 29.61              & 17.77         & 30.46    & 20.79         \\
    Location                    & 0.0933     & -                    & -                & -    & -         \\
    numConnections          & 0.0388     & 158.68           & 164.31         & 215.30    & 173.76         \\
    Subject\_numWords     & 0.0311      & 4.74              & 3.28           & 4.75    & 3.57         \\
    Subject\_isForwarded    & 0.0188     & -                    & -                & -    & -         \\
    Subject\_isReply           & 0.0116     & -                    & -               & -    & -        \\ 
    SummaryNumChars      & 0.0108     & 140.98            & 308.17        & 198.62     & 367.81        \\  
    SummaryRichness        &0.0090     & 0.045               & 0.074          & 0.057      & 0.080        \\ 
    jobLevel                     &0.0088     & 3.41                 & 2.40           & 3.71        & 2.49        \\ \hline
    \end{tabular}
\vspace{5pt}
\caption{Information gain, mean and standard deviation of the 10 most informative features from SPEAR and a combination of BENIGN and SPAM emails. The \emph{subject} features performed best at distinguishing SPEAR emails from non SPEAR emails.}
\label{tab:rankedfeats3}
\end{center}
\end{table}

Contrary to our observations in SPEAR versus SPAM, and SPEAR versus BENIGN emails, we found five \emph{social} features among the top 10 features in this analysis. These were the \emph{Location, numConnections, SummaryNumChars, Richness}, and \emph{jobLevel} features. Although there was a significant difference between the average number of LinkedIn connections in the two datasets, this feature did not have much information gain associated with it due to the very large standard deviation.

\section{Discussion} \label{sec:conclusion}

In this paper, we attempted to utilize \emph{social} features from LinkedIn profiles of employees from 14 organizations, to distinguish between spear phishing and non spear phishing emails. We extracted LinkedIn profiles of 2,434 employees who received a 4,742 targeted spear phishing emails; 5,914 employees who received 9,353 spam or phishing emails; and 1,240 Enron employees who received 6,601 benign emails. 
We performed our analysis on a real world dataset from Symantec's enterprise email scanning service, which is one of the biggest email scanning services used in the corporate organizational level. Furthermore, we targeted our analysis completely on corporate employees from 14 multinational organizations instead of random real-world users. The importance of studying spear phishing emails in particular, instead of general phishing emails, has been clearly highlighted by Jagatic et al.~\cite{Jagatic2007}. 
We performed three classification tasks viz. spear phishing emails versus spam / phishing emails, spear phishing emails versus benign emails from Enron, and spear phishing emails versus a mixture of spam / phishing emails and benign Enron emails. We found that in two out of the three cases, social features extracted from LinkedIn profiles of employees did not help in determining whether an email received by them was a spear phishing email or not. Classification results from a combination of spam / phishing, and benign emails showed some promise, where \emph{social} features were found to be slightly helpful. The depleted \emph{email} feature sets in this case, however, aided the enhancement in classifier performance. 
We believe that it is safe to conclude that publicly available content on an employee's LinkedIn profile was not used to send her targeted spear phishing emails in our dataset. However, we cannot rule out the possibility of such an attack outside our dataset, or in future. These attacks may be better detected with access to richer \emph{social} features. This methodology of detecting spear phishing can be helpful for safeguarding soft targets for phishers, i.e. those who have strong social media footprint. Existing phishing email filters and products can also exploit this technique to improve their performance, and provide personalized phishing filters to individuals.

There can be multiple reasons for our results being non-intuitive. Firstly, the amount of social information we were able to gather from LinkedIn, was very limited. These limitations have been discussed in Section~\ref{sec:linkedin_data}. It is likely that in a real-world scenario, an attacker may be able to gain much more information about a victim prior to the attack. This could include looking for the victim's profile on other social networks like Facebook, Twitter etc., looking for the victim's presence on the Internet in general, using search engines (Google, Bing etc.), and profiling websites like Pipl~\footnote{\url{https://pipl.com/}}, Yasni~\footnote{\url{http://www.yasni.com/}} etc. The process of data collection by automating this behavior was a time consuming process, and we were not able to take this approach due to time constraints. Secondly, it was not clear that which all aspect(s) of a user's social profiles were most likely to be used by attackers against them. We tried to use all the features viz. textual information (summary and headline), connectivity (number of connections), work information (job level, and job type) and location information, which were made available by LinkedIn API, to perform our classification tasks. However, it is possible that none of these features were used by attackers to target their victims. In fact, we have no way to verify that the spear phishing emails in our dataset were even crafted using features from social profiles of the victims. These reasons, however, only help us in better understanding the concept of using social features in spear phishing emails.

In terms of research contributions, this work is based on a rich, true positive, real world dataset of spear phishing, spam, and phishing emails, which is not publicly available. We believe that characterization of this data can be very useful for the entire research community to better understand the state-of-the-art spear phishing emails that have been circulated on the Internet over the past two years. To maintain anonymity and confidentiality, we could not characterize this data further, and had to anonymize the names of the 14 organizations we studied. Also, after multiple reports highlighting and warning about social media features being used in spear phishing, there does not exist much work in the research community which studies this phenomenon.

We would like to emphasize that the aim of this work is not to try and improve the existing state-of-the-art phishing email detection techniques based on their header, and content features, but to see if the introduction of social media profile features can help existing techniques to better detect spear phishing emails. We believe that this work can be a first step towards exploring threats posed by the enormous amount of contextual information about individuals, that is present on online social media. In future, we would like to carry out a similar analysis using the same email dataset, with more social features, which we were not able to collect in this attempt due to time constraints. We would also like to apply more machine learning and classification techniques like Support Vector Machines, Stochastic Gradient boosting techniques etc. on this dataset to get more insights into why social features did not perform well.


\section{Acknowledgement}
We would like to thank the Symantec team for providing us with the email data that we used for this work. We would also like to thank the members of Precog Research Group, and Cybersecurity Education and Research Center at IIIT-D for their support.

\bibliographystyle{abbrv}
\bibliography{bib}

\end{document}